\documentclass[12pt,preprint]{aastex}

\usepackage{emulateapj5}
\usepackage{apjfonts}

\begin{document}
\def\gapprox{\mathrel{\vcenter{\offinterlineskip \hbox{$>$}
    \kern 0.3ex \hbox{$\sim$}}}}
\def\lapprox{\mathrel{\vcenter{\offinterlineskip \hbox{$<$}
    \kern 0.3ex \hbox{$\sim$}}}}

\title{The Magnetic Rayleigh-Taylor Instability in Three Dimensions}

\author{James M. Stone and Thomas Gardiner\altaffilmark{1}}
\affil{Department of Astrophysical Sciences, Princeton University, Princeton,
NJ 08544}
\altaffiltext{1}{Present address: 3915 Rayado Pl NW, Albuquerque, NM 87114}

\begin{abstract}
We study the magnetic Rayleigh-Taylor instability in three dimensions,
with focus on the nonlinear structure and evolution that results from
different initial field configurations.  We study strong fields in the
sense that the critical wavelength $\lambda_c$ at which perturbations
along the field are stable is a large fraction of the size of the
computational domain.  We consider magnetic fields which are initially
parallel to the interface, but have a variety of configurations, including
uniform everywhere, uniform in the light fluid only, and fields which
change direction at the interface.  Strong magnetic fields do not suppress
instability, in fact by inhibiting secondary shear instabilities, they
reduce mixing between the heavy and light fluid, and cause the rate of
growth of bubbles and fingers to increase in comparison to hydrodynamics.
Fields parallel to, but whose direction changes at, the interface produce
long, isolated fingers separated by the critical wavelength $\lambda_c$,
which may be relevant to the morphology of the optical filaments in the
Crab nebula.

\end{abstract}


\section{Introduction}

There are a number of astrophysical systems in which the magnetic
Rayleigh-Taylor instability (RTI) is expected to be important, for
example accretion onto magnetized compact objects (Arons \& Lea 1976;
Wang \& Nepveu 1983; Wang, Nepveu, \& Robertson 1984), buoyant bubbles
generated by radio jets in clusters of galaxies (Robinson et al. 2005;
Jones \& De Young 2005; Ruszkowski et al. 2007)), the emergence of
magnetic flux from the solar interior and the formation of flux tubes
(Isobe et al. 2005; 2006; and references therein), and at both the contact
discontinuity between the shocked circumstellar medium and ejecta in
supernovae remnants (Jun \& Norman 1996a; b), and in the thin shell of
ejecta swept up by a pulsar wind (Hester et al. 1996, hereafter H96;
Bucciantini et al.  2004).  For the idealized case of two inviscid,
perfectly conducting fluids separated by a contact discontinuity with
a uniform magnetic field ${\bf B}$ parallel to the interface undergoing
constant acceleration $g$, then a linear analysis (Chandrasekhar 1961)
demonstrates that for modes parallel to the magnetic field there is a
critical wavelength
\begin{equation}
\lambda_c = \frac{B^2}{g(\rho_h - \rho_l)}
\end{equation}
below which instability is completely suppressed, where $\rho_h$ and
$\rho_l$ are the densities in the heavy and light fluids respectively,
and we have chosen a system of units in which the magnetic permeability
$\mu=1$.  At larger wavelengths, the growth rate is reduced compared
to the hydrodynamic case, and there is a peak growth rate occurring at
a wavelength $\lambda_{\rm max} = 2\lambda_c$.  Equation (1) can also
be thought of as a condition on the magnetic field: instability on a
scale $L$ parallel to the field requires $B < B_c \equiv [Lg(\rho_h -
\rho_l)]^{1/2}$.  Modes perpendicular to the field are unaffected,
and have the same growth rate and stability condition as in pure
hydrodynamics.  The highly anisotropic nature of the growth rate of
modes parallel versus perpendicular to the field suggests that it is
important to study the nonlinear regime of the magnetic RTI in full
three dimensions.

One of the most compelling applications of the magnetic RTI is to the
structure of the optical filaments in the Crab nebula (H96).  As the
low density, highly magnetized synchrotron nebula powered by the Crab
pulsar sweeps up the stellar ejecta, the interface between the two
is RT unstable, resulting in radially orientated filaments that point
to the center of the synchrotron nebula.  H96 have suggested the long,
widely spaced filaments observed by HST are a consequence of suppression
of short wavelength modes due to the magnetic field in the synchrotron
plasma, since the filaments bear no resemblance to the turbulent mixing
layer that results from the RTI in hydrodynamics (Dimonte et al 2004,
hereafter D04), but are a better fit to the morphology that results from
the magnetic RTI in two-dimensions (Jun, Norman, \& Stone 1995).  However,
in purely hydrodynamic simulations of the RTI in the spherically expanding
shell swept up by the pulsar wind, Jun (1998) was able to reproduce the
morphology and separation of the fingers remarkably well, suggesting
than geometrical effects might be important.  Since the simulations were
performed in two-dimensions assuming axial symmetry, it is unclear if
the isolated fingers will persist in three-dimensional hydrodynamics, or
whether MHD effects are indeed essential.  More recently, Bucciantini
et al. (2004) have presented the most realistic numerical models
of the filaments in the Crab nebula to date, using two-dimensional
MHD simulations of the expanding shell and nebula.  In their
more realistic treatment of the conditions at the unstable interface,
they find fields near equipartition can completely suppress the RTI.
However, as they point out, because of the anisotropic nature of the
magnetic RTI, three-dimensional effects are critical and need to be
included in future studies.

Since fully three-dimensional MHD simulations in a spherical geometry
which can follow the expanding shell of ejecta are computationally challenging,
it is worthwhile to begin investigation of three-dimensional effects
in the idealized plane parallel case.  Recently, we have described an
extensive study of the nonlinear evolution of the magnetic RTI in a
three-dimensional planar geometry, focusing on the effect of varying the
field strength on the growth rate of fingers and bubbles at the interface,
and on the amount of mixing between the heavy and light fluids (Stone
\& Gardiner 2007, hereafter Paper I).  To facilitate comparison with
previous experimental and computational studies of the hydrodynamic RTI
(D04), a relatively modest difference in density between the fluids
was chosen, that is $\rho_h/\rho_l = 3$.  In this paper, we extend the
study by considering a more astrophysically relevant density ratio,
$\rho_h/\rho_l = 10$, and by focusing on the effect of strong magnetic
fields (in the sense that $\lambda_c \sim L$, where L is the size of the
computational domain) on the suppression of the RTI in three dimensions.

A number of studies of magnetic buoyancy instabilities in three
dimensions have been reported, both in the context of the emergence
of new magnetic flux from the solar photosphere (Wissink et al. 2000;
Fan 2001; Isobe et al. 2005; 2006), and the nonlinear evolution of the
Parker instability in the galactic disk (Kim, Ostriker, \& Stone 2002;
Kosi\'{n}ski \& Hanasz 2007).  In these studies, the magnetic field is strong
enough for the ratio of thermal to magnetic pressure $\beta \sim 1$, so
that the magnetic field not only plays a significant role in the support
of the initial equilibrium state, but also is responsible for driving
buoyant motions.  In contrast, we study weak fields in the sense that
$\beta \gg 1$, so that the magnetic field plays almost no role in the
vertical equilibrium, and the RTI is driven by the buoyancy of the fluid.
Our goal is to study how magnetic fields affect the evolution
of the classical RTI.

Our primary conclusions are that in three dimensions, uniform magnetic
fields do not suppress the RTI due to the growth of interchange modes
perpendicular to the field.  In fact, since magnetic fields suppress
secondary Kelvin-Helmholtz instabilities and therefore mixing between
the heavy and light fluids, the growth rate of bubbles and fingers is
in fact enhanced in the magnetic RTI compared to the hydrodynamic case.
We explore a variety of initial field
configurations, including uniform fields, uniform fields in the light
fluid only, and fields with a rotation at the interface, and we show that
well separated, long fingers reminiscent of the optical filaments in
the Crab nebula can be generated if the magnetic field direction changes
through large angles over a distance short compared to $\lambda_c$.

\section{Method}
We solve the equations of ideal MHD with a constant vertical
acceleration ${\bf g} = (0,0,g)$ 
\begin{eqnarray}
\frac{\partial \rho}{\partial t} +
{\bf\nabla\cdot} \left(\rho{\bf v}\right) & = & 0
\label{eq:cons_mass} \\
\frac{\partial \rho {\bf v}}{\partial t} +
{\bf\nabla\cdot} \left(\rho{\bf vv} - {\bf BB}\right) +
{\bf \nabla} P^{*} & = & \rho {\bf g}  \\
\frac{\partial {\bf B}}{\partial t} +
{\bf\nabla}\times \left({\bf v} \times {\bf B}\right) & = & 0 \\
\frac{\partial E}{\partial t} +
\nabla\cdot((E + P^*) {\bf v} - {\bf B} ({\bf B \cdot v})) & = & \rho {\bf v} \cdot {\bf g}
\label{eq:cons_energy}
\end{eqnarray}
The total pressure $P^* \equiv P + ({\bf B \cdot B})/2$,
where $P$ is the gas pressure.  The total energy density $E$ is 
\begin{equation}
E \equiv \epsilon + \rho({\bf v \cdot v})/2 + ({\bf B \cdot B})/2 ~.
\end{equation}
where $\epsilon$ is the internal energy density.  We use an ideal gas
equation of state for which $P = (\gamma - 1) \epsilon$, where $\gamma$
is the ratio of specific heats.  We use $\gamma=5/3$ in this paper.
In relativistic plasmas such as synchrotron nebulae $\gamma=4/3$ would
be more appropriate.  However, given our choice for the numerical value
of $g$ and the size of the computational domain (see below), the flows
induced by the magnetic RTI are subsonic and nearly incompressible.
Thus, varying the adiabatic index should have little effect on the
results reported here.

The three-dimensional computational domain is of size $L
\times L \times 2L$, where $L=0.1$.
Periodic boundary conditions are used in the transverse
($x-$ and $y-$) directions, and reflecting boundary conditions are used at
the top and bottom.  The origin of the $z-$coordinate
is centered in the domain, so that the computations span $-0.1 \leq z
\leq 0.1$  The upper half of the domain ($z>0$) is filled with heavy
fluid of density $\rho_h=10$, while in the lower half ($z<0$) the density
of the light fluid is $\rho_l=1$.  Thus, the Atwood number
\begin{equation}
 A \equiv \frac{\rho_h - \rho_l}{\rho_h + \rho_l} = \frac{9}{11}.
\end{equation}
Most of the experimental studies of the hydrodynamic RTI used
to validate computational methods (D04) use $A=1/2$.  In Paper I we
studied the magnetic RTI with $A=1/2$;
in this paper we study the high Atwood number regime which is more
relevant to most astrophysical systems.

Initially the gas is in magnetohydrostatic equilibrium, with the amplitude
of the gas pressure chosen so that the sound speed in the light fluid
$c_s=1$ at the interface, thus
\begin{equation}
P^*(z) = \frac{3}{5} - g \rho z + B^{2}/2
\end{equation}
The sound crossing time in the light fluid at the interface $t_s =0.1$.
We choose $g=0.1$, thus the ratio of the free-fall velocity to the sound
speed $\sqrt{gL}/c_s = 0.1$, implying the induced flows should be
nearly incompressible.

The magnetic field is initialized with an amplitude $B_{0}$ that is chosen
to be a fixed fraction of the critical field strength $B_c$ at which
there are no unstable modes within $L$, we choose $B_{0} \approx 0.6B_c$.
From equation (1), the critical
wavelength at which all modes are suppressed $\lambda_c/L \approx 0.35$.
The field is always initially parallel to the interface, but has a variety
of different initial configurations which will be described along with the
results of each individual simulation in \S 3.  The ratio of the gas to
magnetic pressure at the interface $\beta = 480$ in all the runs.  Thus,
although we study strong fields in the sense that $\lambda_c \sim L$,
the energy density in the field is far below equipartition, and the field
plays little role in the initial vertical equilibrium.  Increasing the
size of the computational domain $L$ to accommodate a larger $\lambda_c$
associated with stronger, near-equipartition fields, or simply lowering
the sound speed to decrease $\beta$ in the present simulations, will
both produce flows in which $\sqrt{gL}/c_s$ is increased, and therefore
are more compressible.

To seed the RTI, zone-to-zone perturbations are added to the vertical
velocity $v_z$ throughout the volume with an amplitude $A$ that is kept
small compared to the sound speed, and is decreased toward the vertical
boundaries; thus $A = A_0 R (1 + \cos{2\pi z/L})$ where $A_0=0.005$, and
$R$ is a random number between -1 and 1.  The maximum perturbed velocity
is only 1\% of the sound speed in the light fluid at the interface.

The computations presented in this paper use Athena, a new MHD code
that implements a recently developed
Godunov method for compressible MHD (Gardiner \& Stone 2005; 2007).
A complete description of the algorithm, including the results of an
extensive series of test problems, is given in these references.  All of
the simulations use a grid of $256 \times 256 \times 512$, which means the
critical wavelength $\lambda_c$ is resolved with nearly 100 grid points.
Our numerical resolution is much higher than used in previous work
(for example, the few 3D simulations reported
in Jun, Norman \& Stone 1995), and uses stronger initial fields.

In Paper I, we presented a comprehensive convergence study of our
numerical algorithms for the magnetic RTI in two dimensions, focused
on the amount of mass mixing due to numerical effects.  For single
mode perturbations, features such as the shape and height of the
interface at a fixed time were converged with 32 or more grid points
per wavelength.  The amount of mixing between heavy and light fluids
was also found to converge to zero at first order, independent of the
magnetic field strength.  First order convergence is consistent with
mixing being proportional to the width of the interface between the
heavy and light fluids (which cannot be smaller than one grid cell).
With multimode perturbations, the degree of mixing does not converge to
zero, because at higher resolution there are more small scale distortions
in the interface which increase its surface area.  Convergence of
the mixing to zero with multimode perturbations therefore requires
the introduction of surface tension or viscosity to create a fixed
small scale below which the interface is smooth.  Instead, in this
paper we compute all solutions at the highest resolution we can afford
(so that they are all at the same, high Reynolds and magnetic Reynolds
numbers), and focus on the {\em comparison} of features with different
field strengths and geometries that occur at these Reynolds numbers.
In this way, we can isolate the effects of changing field strength or
geometry from the effect of changing the numerical diffusion.

\section{Results}

We describe the results from simulations that use a variety of different
initial magnetic field configurations. 

\subsection{Uniform Field versus Hydrodynamics}

We begin with the evolution in a uniform magnetic
field parallel to the interface and along the $x-$axis, ${\bf B} =
(B_{0},0,0)$.  For comparison purposes, we also describe the results of
a hydrodynamical calculation, computed with the same parameters, grid,
and numerical algorithm.  Hereafter, we refer to the uniform field case
as run U, and the hydrodynamical simulation as run H.

Figure 1 shows isosurfaces of the density, along with slices of the
density at the edges of the computational domain, at two times during
the evolution for both runs H and U.  The hydrodynamic case shows
the typical evolution of the RTI into a turbulent mixing layer (D04).
In hydrodynamics, short wavelength modes grow fastest, thus at early times the
instability is dominated by bubbles and fingers at small scales.
Secondary Kelvin-Helmholtz instabilities, associated with the shear
between the rising and descending plumes, give the tips a ``mushroom-cap"
appearance, and cause some of the fingers to break up.  At late times,
mergers between bubbles favors growth of structure at larger scales, while
secondary instabilities continue to distort the plumes and cause mixing.
Note the large fraction of fluid at intermediate densities (green colors)
at late times in the hydrodynamic case.

In the uniformly magnetized simulation run U, the early nonlinear phase
of the RTI shows the strongly anisotropic structure of modes introduced by
the magnetic field.  Perpendicular to the field (along the $y-$axis), 
interchange modes grow fastest at short wavelength, whereas along the field
short wavelengths are suppressed.  As a result, the interface develops a
filamentary structure that is strongly reminiscent of the structure
reported by Isobe et al (2005; 2006) in simulations of flux tubes emerging from
the solar photosphere.  At late times, fluid flowing along flux tubes
collects at bubbles and fingers at the tips (similar to the nonlinear
evolution of the Parker instability, Kim et al. 1998), which are then
wrinkled by interchange instability at their surface.  This produces
large-scale smooth bubbles.  Slices along the edges of the domain reveal
far less mixing than in the hydrodynamic case.  Note in three-dimensions
the magnetic RTI in a uniform field does not result in isolated, long
fingers comparable to the observations of the Crab (H96).

One measure of the rate of growth of the RTI is the time evolution of the
height $h$ of bubbles from the interface.  Self-similar arguments (D04)
predict that
\begin{equation}
h = \alpha Agt^2
\end{equation}
where $\alpha$ is a dimensionless constant.  The experimentally measured
value is $\alpha = 0.057 \pm 0.008$.  Without specialized
front-tracking algorithms that can prevent mixing between the fluids
at the grid scale, most numerical methods give a value for $\alpha$ which
is about a factor of two smaller (D04, Paper I).

Figure 2 plots the location of the tips of the rising
bubbles as a function of time in both runs H and U.  At any instant in
time, we define the vertical location of the tips of the fingers as the
point where the horizontally averaged fraction of the heavy fluid
\begin{equation}
 \langle f_h \rangle = \int_x \int_y f_h dx dy/L^2
\end{equation}
is 0.95, where for incompressible fluid with $\rho_h=10$ and $\rho_l=1$
the fraction of heavy fluid in any cell is $f_h = (\rho -1)/9$.
(To account for the effects of compressibility, we choose $f_h=0.95$
rather than one to mark the boundary of the mixing region.)  From figure
2, we see that after an initial rise, the increase in $h$ in both
hydrodynamics and MHD follows the expected self-similar scaling equation
(9).  In hydrodynamics, the slope $\alpha =0.03$, whereas in MHD the
slope $\alpha =0.05$ where we have ignored the final point in both cases,
since it is undoubtedly affected by the reflecting boundary conditions
at the top of the domain $h/L=1$.  It is clear that the bubbles rise
{\em faster} in MHD than in the hydrodynamic RTI, in agreement with the
results at $A=1/2$ (Paper I).  As discussed in \S 3.5, this is primarily
due to the reduction of mixing in the MHD case.

\subsection{Field in Light Fluid Only}

In the magnetic RTI associated with some astrophysical systems, such
as the interface between the pulsar wind nebula and the supernova
ejecta in the Crab nebula, only the light fluid is expected to be
strongly magnetized.  Given that the results in section \S 3.1 show
that strong, uniform fields do not suppress the RTI, it is
unlikely that a strong field in the light fluid only will inhibit
instability.  Nonetheless, it is of interest to investigate the
structure of the nonlinear regime in this case.

Figure 3 plots isosurfaces of the density, along with slices of the
density at the edges of the computational domain, at two times during
the evolution of a simulation in which the magnetic field is uniform,
parallel to the interface, and along the $x-$axis, ${\bf B} = (B_{0},0,0)$
in the light ($\rho=1$) fluid only, with ${\bf B} = 0$ everywhere else.
As before, we choose $B_0 = 0.6 B_{c}$.
The gas pressure is increased in the heavy fluid above the interface
so that the total pressure is continuous, that is exact vertical
equilibrium is maintained initially.  We refer to this calculation as
run LF hereafter.

It is instructive to compare the structures observed in figure 3 with
the uniform field case (bottom row of figure 1).  At the early time in
run LF, the fingers and bubbles are not elongated along the field as
in run U.  Instead, the structure is nearly isotropic, similar to the
hydrodynamic case but with less small scale structure.  At late time,
large smooth bubbles emerge in run LF that appear isotropic.  Overall,
the three-dimensional structure of the fingers and bubbles in run LF is
intermediate between the hydrodynamic and uniformly magnetized runs.
The density slices at the edge of the domain show much less mixing than
run H.  The height of the bubbles and degree of mixing (revealed by the
density slices at the edge of the domain) show much more similarity to
run U; these will be analyzed further in \S 3.5.  Once again, we see that
in three-dimensions, strong uniform fields in the light fluid
are unable to inhibit the RTI.

\subsection{Fields with a Discontinuous Rotation}

In most astrophysical systems, there is no reason to expect the
magnetic field has the same geometry in both the light and heavy fluids.
Since only unstable modes parallel to the magnetic field are suppressed,
rotating the field near the interface will inhibit
modes in multidimensions.
To investigate this regime we have performed simulations in
which the magnetic field is rotated discontinuously through large angles
at the interface.
In the first simulation, hereafter referred to as run R45, the field is
rotated through $45^{\circ}$, that is ${\bf B} = (B_{0},0,0)$ in the
light fluid ($z<0$), and ${\bf B} = (B_{0}/\sqrt{2},B_0/\sqrt{2},0)$
in the heavy fluid ($z>0$).  In the second simulation, hereafter referred
to as run R90, the field is rotated through $90^{\circ}$, that is ${\bf
B} = (B_{0},0,0)$ in the light fluid, and ${\bf B} = (0,B_0,0)$ in the
heavy fluid.  In both cases, there is a current sheet at the
interface.

Figure 4 plots isosurfaces of the density, along with slices of the
density at the edges of the computational domain, at two times during
the evolution of both runs R45 and R90.  At early times in both cases,
filamentary structures appear at an angle roughly half-way between the
direction of the field in the heavy and light fluids (about $22^{\circ}$
with respect to the $x-$axis in R45, $45^{\circ}$ with respect to
the $x-$axis in R90), most likely because the magnetic tension forces
which are proportional to ${\bf k}\cdot{\bf B}$ are minimized in this
direction.  Analysis of the magnetic field and velocity perturbations
at this time shows flow occurs along the field lines into the ridges.
Pure interchange modes are no longer possible with rotated fields,
and the growth of perturbations requires ${\bf k}\cdot{\bf B} \ne 0$ in
either the light or heavy fluids, or both.  Note the amplitude of perturbations
is much smaller in R90 at early times in comparison to R45, and only
long wavelength modes are present.

By $t/t_s=40$, the interface in both R45 and R90 is strongly distorted
by RTI.  Interestingly, the structure of modes at late times is quite
different from previous cases.  Isolated, large scale bubbles dominate,
with very smooth surfaces, and bulbous tips.  The spacing between bubbles
is roughly the critical wavelength $\lambda_c$.  The structure of R90 is
particularly interesting.  The fingers in this case are nearly isotropic,
and have a length which significantly exceeds $\lambda_c$.  The surface
of the bubbles is extremely smooth, whereas in R45 there is some evidence
for wrinkling due to interchange modes.  The interface between the light
and heavy fluids is remarkably thin in R90.  At the faces of the volume,
the density slices reveal very little material at densities intermediate
to the values of the isosurfaces (at $\rho=9.9$ and 1.1 respectively).
Thus, the faces of the volume are transparent, and the interior of the
bubbles is clearly visible.  Contrast this to run H in figure 1, where
the slice at the edge of the domain revealed a turbulent mixing layer.
A more quantitative analysis of mixing in all the runs
will be presented in \S 3.5.

\subsection{Fields with Continuous Rotation}

In the previous section, the direction of the magnetic field was changed
discontinuously at the interface, resulting in a current sheet.  It is
possible that in many astrophysical systems, the direction of the field
varies smoothly on many different scales.  To investigate the effect
this might have on the magnetic RTI, we consider the case where the field
amplitude is constant everywhere, while the direction is rotated through
a large angle (we choose $90^{\circ}$) over a finite vertical distance
$L_{rot}$. More specifically, for $z<-L_{rot}/2$ the field is ${\bf B} =
(B_{0},0,0)$, for $-L_{rot}/2<z<L_{rot}/2$ the direction of the field
varies linearly with $z$ from along the $x-$axis to along the $y-$axis
while the amplitude is fixed at $B_{0}$, and for $z>L_{rot}/2$ the field
is ${\bf B} = (0,B_{0},0)$.  Note this geomtry results in a current
layer with constant amplitude in the region $-L_{rot}/2<z<L_{rot}/2$.
If $L_{rot} \ll \lambda_{c}$ we expect this initial configuration to
evolve similar to the discontinous rotation run R90 studied in \S3.3,
while if $L_{rot} \gg \lambda_{c}$ it will evolve like the uniform field
case run U studied in \S3.1.  Here we choose $L_{rot}/\lambda_{c} =
0.5$, and hereafter refer to this calculation as run C90.

In fact, we find at late times the structure that emerges from the
magnetic RTI in run C90 is remarkably similar to that produced in run R90.
For example, at $t/t_s=40$s, isolated smooth bubbles are produced with
similar sizes and spacing as observed in figure 4.  Conversely, we
find at early times there is little suppression of interchange modes.
This is not suprising: the fastest growing modes occur at the largest
wavenumbers, and therefore have wavelengths much smaller than $L_{rot}$.
On these scales, the early evolution of the interface is as if the field were
uniform (run U).
Figure 5 plots the height of bubbles in run C90 versus the uniform field
case run U.  The evolution of both is very similar.  Our results confirm
the intuition that changes in the direction of the field at the interface
must be on very small scales to inhibit the interchange modes.

\subsection{Mixing}

The amount of mixing between the heavy and light fluids strongly affects
the rate at which bubbles and fingers are displaced from the interface
(D04, Paper I).  The presence of even a weak field can, through the action
of tension forces at small scales, significantly reduce mixing in 
comparison to hydrodynamics (Paper I).  Here we investigate mixing in the
simulations presented above.

Figure 6 plots the height of bubbles above the interface, defined using
the point at which $\langle f_h \rangle=0.95$, for runs U, LF, R45
and R90.  In every case, at late times the height $h$ grows as $t^2$,
as expected (equation 9).  However, in R45, and especially R90, growth
is delayed.  The slope of the lines, as measured by the dimensionless
constant $\alpha$ are remarkably similar, $\alpha = 0.050 \pm 0.005$.
The decrease in the slope at late time in each model is most likely an
influence of the upper (reflecting) boundary condition rather than a
divergence from the self-similar evolution.

It is useful to define a mixing parameter $\Theta$ as
\begin{equation}
\Theta =4\langle f_h f_l \rangle
\end{equation}
The peak value of $\Theta$ is one, and occurs when $f_h=f_l=1/2$, that
is in regions that are fully mixed.  In regions that are not mixed,
$\Theta=0$.  Figure 7 plots the profile of $\Theta$ versus height in
runs H, U, and R90.  Note in the hydrodynamic case run H, the mixing
parameter is close to the theoretical maximum near the original location
of the interface $z=0$.  This quantifies the result which is evident from
a visual inspection of figure 1, namely the hydrodynamic RTI results in
a turbulent mixing zone which is dominated by material at intermediate
densities.  On the other hand, the uniformly magnetized case run U
shows far less mixing than run H, again a fact which is evident from
the lower panels of figure 1.  Finally, run R90 shows the least mixing,
with a peak value of $\Theta$ which is five times smaller than the peak
value in run H.  At the peak of $\Theta$ in run R90, the horizontally
averaged fraction of heavy fluid $\langle f_h \rangle = 0.2$, indicating
the fingers of heavy fluid occupy a much smaller volume than the bubbles
of light fluid.  Again, all of these results are evident from figure 4,
where density slices at the edge of the domain show the mixing layer
between the two fluids to be very thin, and that the bubbles of light
fluid fill most of the volume.

\subsection{Magnetic Field Evolution}

Self-similar arguments predict that the rate of growth of the height $h$
of bubbles and fingers should be proportional to $t^2$ (equation 9).
Since the amount
of gravitational binding energy released by the descending plumes of heavy
fluid is proportional to $h^2$ (the energy released is the product of
the mass involved in the flow and the distance it falls, both of which
are proportional to $h$), we expect the rate of growth of
the kinetic and magnetic energies in the RTI should be proportional
to $t^{4}$.

Figure 8 plots each component of the volume averaged
kinetic and magnetic energies, normalized by the initial volume averaged
magnetic energy $B_{0}^{2}/2$, in runs U and R90 versus $t^4$.  The magnetic energy
associated with the horizontal components of the field have their initial
values subtracted as appropriate, thus the plot shows the fractional change
in the energies.  Note that at early times, the curves are straight lines,
indicating the expected scaling with $t^4$ is recovered.  In each
case the vertical components of the energies dominate, and in the
horizontal directions there is rough equipartition between kinetic and
magnetic energies.  The kinetic energy associated with the $y-$component
of the kinetic energy is larger in comparison to the $x-$component
in run U since motions perpendicular to the field are favored by
interchange modes, which we have shown are important in strong uniform
fields.   The amplification of the vertical field is larger in run R90
in comparison to run U, although the total magnetic energy in all 
components of the field is roughly the same at late times in both cases.
This is another indication that run R90 leads to ordered, vertical flows
and columns, whereas larger amplitude horizontal flows (and therefore
more mixing and less ordered fingers) are produced in uniform fields.
In both cases the magnetic RTI leads to significant amplification of
magnetic energy.

It is worth emphasizing that the time evolution of volume averaged
quantities as shown in figure 8 is controlled by a number of dimensionless
parameters, including the ratio of the critical wavelength to the size
of the computational domain $\lambda_c/L$ and the ratio of the free fall
to the sound crossing time $\sqrt{\lambda_c/g}/t_{s}$.  We have studied
strong fields in the sense that $\lambda_c/L \sim 1$.  If the calculations
were repeated with identical parameters but in a much larger domain,
then the evolution would resemble the weak field simulations presented
in Paper I.  That is, if the calculations presented here were continued in
a much larger domain, so that the height of the fingers and bubbles $h
\gg \lambda_c$, then the flow would become more hydrodynamic, a turbulent
mixing zone would emerge, and once the vertical field is a large fraction
of the initial horizontal value, the $t^4$ scaling of energies is broken
(Paper I).

\section{Summary and Discussion}

We have shown that strong, uniform magnetic fields cannot suppress the
RTI in three dimensions.  In the linear regime only long wavelength modes
parallel to the magnetic
field are unstable; interchange modes perpendicular to the
field are unaffected, and grow at the same rate as in
hydrodynamics.   We have shown that in the nonlinear regime this leads
to a highly anisotropic structure.  At late times, flow of plasma along
field lines produces large bubbles much as in the Parker instability (Kim
et al 1998), which in turn become wrinkled by secondary interchange modes.

In fact, in one respect strong magnetic fields actually increase
the growth rate of the RTI in the nonlinear regime, in comparison to
hydrodynamics.  Magnetic fields inhibit secondary instabilities and
mixing between the light and heavy fluid.  In turn, the reduction of
mixing causes bubbles (fingers) to rise (fall) more rapidly.  In fact,
the tension force associated with even weak fields can suppress mixing
on small scales, and increase the growth rate of bubbles and fingers
(Paper I).  The suppression of a turbulent mixing layer with even a weak
magnetic field could be relevant to a number of astrophysical systems,
for example the evolution of supernovae remnants (Jun \& Norman 1996a;
b), or the mixing of metals from early generations of stars into the
intergalactic medium.

Although uniform magnetic fields do not suppress the RTI, we have shown
that if the direction of the field changes through a large angle at
the interface, this can delay instability, and significantly alter
the structures that emerge in the nonlinear regime.  We have studied
field geometries that have both discontinuous rotations of the field
at the interface, and continuous rotations over a finite vertical
length $L_{rot}$ at the interface.  When $L_{rot}/\lambda_c \leq 1$,
the nonlinear regime in both these cases is similar, and consists of
isolated, smooth, long fingers and bubbles.

There are several obvious applications of the magnetic RTI to
astrophysical systems.  The first is to the penetration of infalling
plasma into the magnetosphere of an accreting neutron star (Arons \&
Lea 1976; Wang \& Nepveu 1983; Wang, Nepveu, \& Robertson 1984), or to
the confinement of the plasma along field lines at the polar caps (Litwin,
Brown, \& Rosner 2001).  A related problem, confinement of strong vertical
flux tubes at the galactic center, has been investigated by Chandran
(2001).  In each of these cases (except the last), the field is rigidly
anchored at a boundary, whereas we have studied the magnetic RTI with
periodic boundary conditions in both horizontal directions.  The nonlinear
evolution of interchange modes will probably be strongly affected no-slip
boundary conditions on the magnetic field at the edges of the domain,
so our results may only have limited applicability to these systems.

The second is to the stability of buoyant bubbles generated by radio
jets in clusters of galaxies.  Robertson et al. (2004) and Jones \& De Young
(2005) have presented
two-dimensional simulations of the morphology of magnetized, buoyant
bubbles.  However, it is clear that three-dimensional effects will
be very important in this problem, due to the very different behavior
of modes perpendicular versus parallel to the field.  Recent work in 3D
by Ruszkowski et al. (2007) confirms that magnetic fields are unable to
suppress shredding of bubbles in three dimensions unless the coherence length
of the field is larger than the size of the bubble.  In fact,
magnetic fields in cluster gas can alter the dynamics in ways
than go beyond the obvious effects of magnetic stresses.  Due to the long
mean-free-paths of particles, anisotropic heat conduction and viscosity
are important in hot cluster gas.  Balbus (2000) has
shown that the convective stability criterion is fundamentally altered
in a plasma with anisotropic heat conduction (see also Chandran \&
Dennis 2006).  Numerical simulations of the nonlinear regime of this
instability (Parrish \& Stone 2005; 2007) reveal vigorous convective motions
that are quenched only when the plasma becomes isothermal.  Thus, inclusion
of magnetic fields in the dynamics of buoyant bubbles
alters the basic plasma dynamics in ways that warrant
further investigation.

Finally, our results have application to the optical filaments being swept
up by the pulsar wind in the Crab nebula (H96).  It is tempting to compare
the long, well-separated fingers generated with rotated fields (run R90,
figure 4) with the filaments.  However, due to its orientation, figure 4
shows the morphology of the rising bubbles of light fluid, whereas the
observations reveal the morphology of the descending fingers of heavy
fluid.  In figure 9 we plot isosurfaces of the density at $\rho=1.1$,
and slices along the face of the computational domain showing only regions
where $\rho > 1.1$, with the orientation flipped relative to figure 4,
that is with the descending fingers of heavy fluid oriented upward.
Note the long, thin fingers of dense gas along the $y-z$ plane are in
good agreement with the morphology of the fingers in the Crab.
This calculation includes field in both the heavy and light fluids,
although in the Crab only the light fluid (synchrotron nebula) is expected
to be strongly magnetized.  It is an open issue as to whether a strong
field in the light fluid only, whose direction changes on scales small
compared to $\lambda_c$, can reproduce the structures seen in figure 9.
Note that a uniform field in the light fluid only (run LF, see figure 3)
results in structure markedly different than in figure 9. 

Of course, to accurately model the fingers in the Crab nebula, it is
important to include density compression due to cooling, to study
fields near equipartition (which will also increase the importance
of compressibility), and perhaps most importantly, to include the
geometrical effects produced by the spherically expanding shell.
Previous two-dimensional studies have shown that purely hydrodynamical
instability in the appropriate geometry can produce the structure of the
Crab filaments (Jun 1998), and more recently it has shown that strong
fields in this same geometry significantly alter the picture (Bucciantini
et al 2005).  In this paper, we have emphasized the importance of three
dimensional effects on the magnetic RTI.  It will be important to extend
these fully three-dimensional results to the expanding wind geometry
appropriate to the Crab.

\acknowledgements
We thank Jeff Hester for discussions.  Simulations were performed on
the Teragrid cluster at NCSA, the IBM Blue Gene at Princeton University, and
on computational facilities supported by NSF grant AST-0216105.  Financial
support from DoE grant DE-FG52-06NA26217 is acknowledged.

\clearpage

% Figure 1 -- color images of density
\begin{figure}
\epsscale{1.0}
\plotone{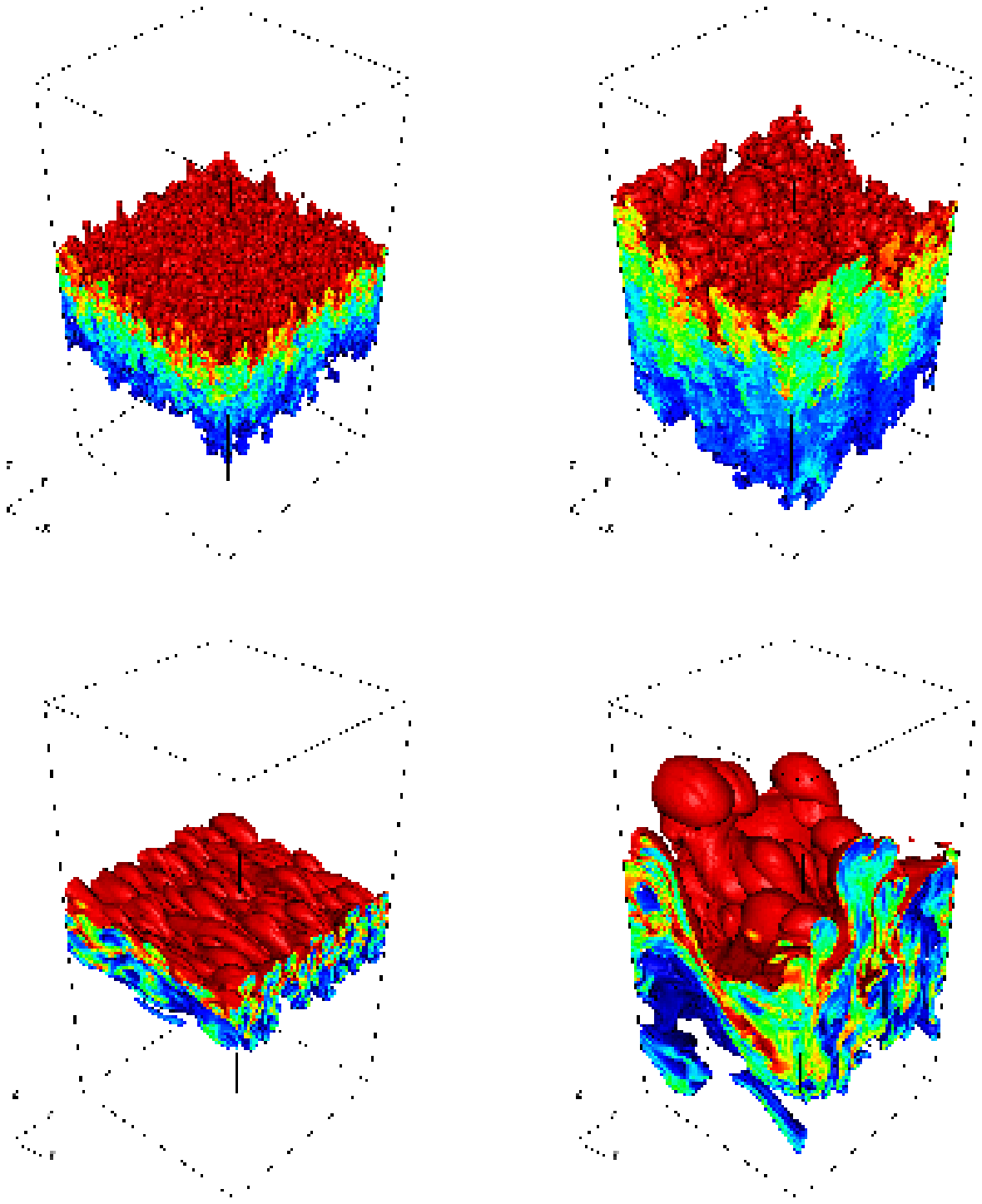}
\figcaption
{Isosurfaces of the density at $\rho=9.9$ and $\rho=1.1$ at times
$t/t_s = 20$ (left panels) and $t/t_s=40$
(right panels) in runs H (top, pure hydro) and
U (bottom, uniform field).  Also shown are slices of the density at the
edges of the computational domain.}
\end{figure}

% Figure 2 -- height vs time
\begin{figure}
\epsscale{0.6}
\plotone{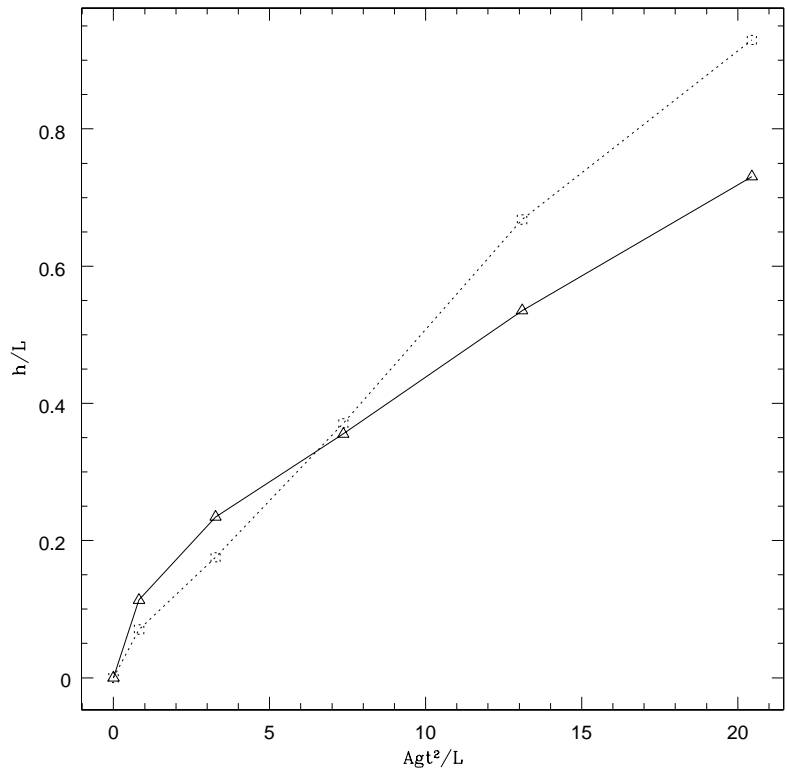}
\figcaption
{Height of bubbles as a function of time in hydrodynamic (run H,
solid line and triangles) and magnetic RTI in a uniform field
(run U, dotted line and squares).  Note bubbles rise {\em faster} in MHD.}
\end{figure}

% Figure 3 -- color images of density, B in light fluid
\begin{figure}
\epsscale{1.0}
\plotone{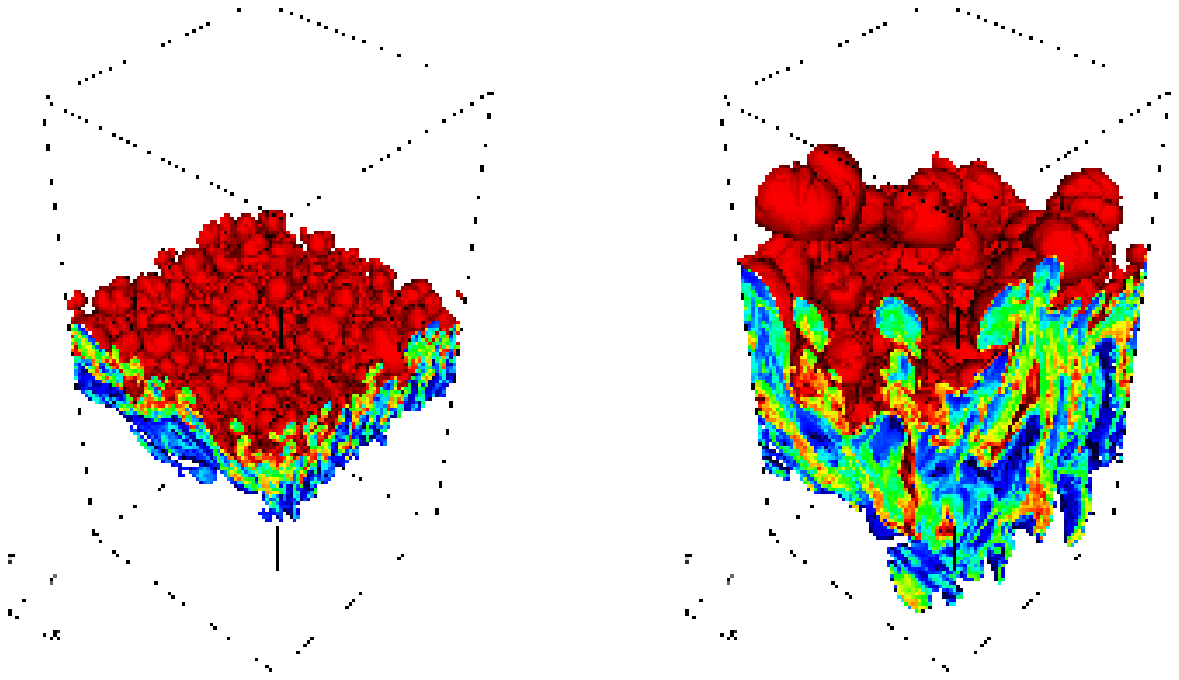}
\figcaption
{Isosurfaces of the density at $\rho=9.9$ and $\rho=1.1$ at times $t/t_s =
20$ (left panel) and $t/t_s=40$ (right panel) in run LF (uniform field
in light fluid only).  Also shown are slices of the density at the edges
of the computational domain.}
\end{figure}

% Figure 4 -- color images of density, rotated B
\begin{figure}
\epsscale{1.0}
\plotone{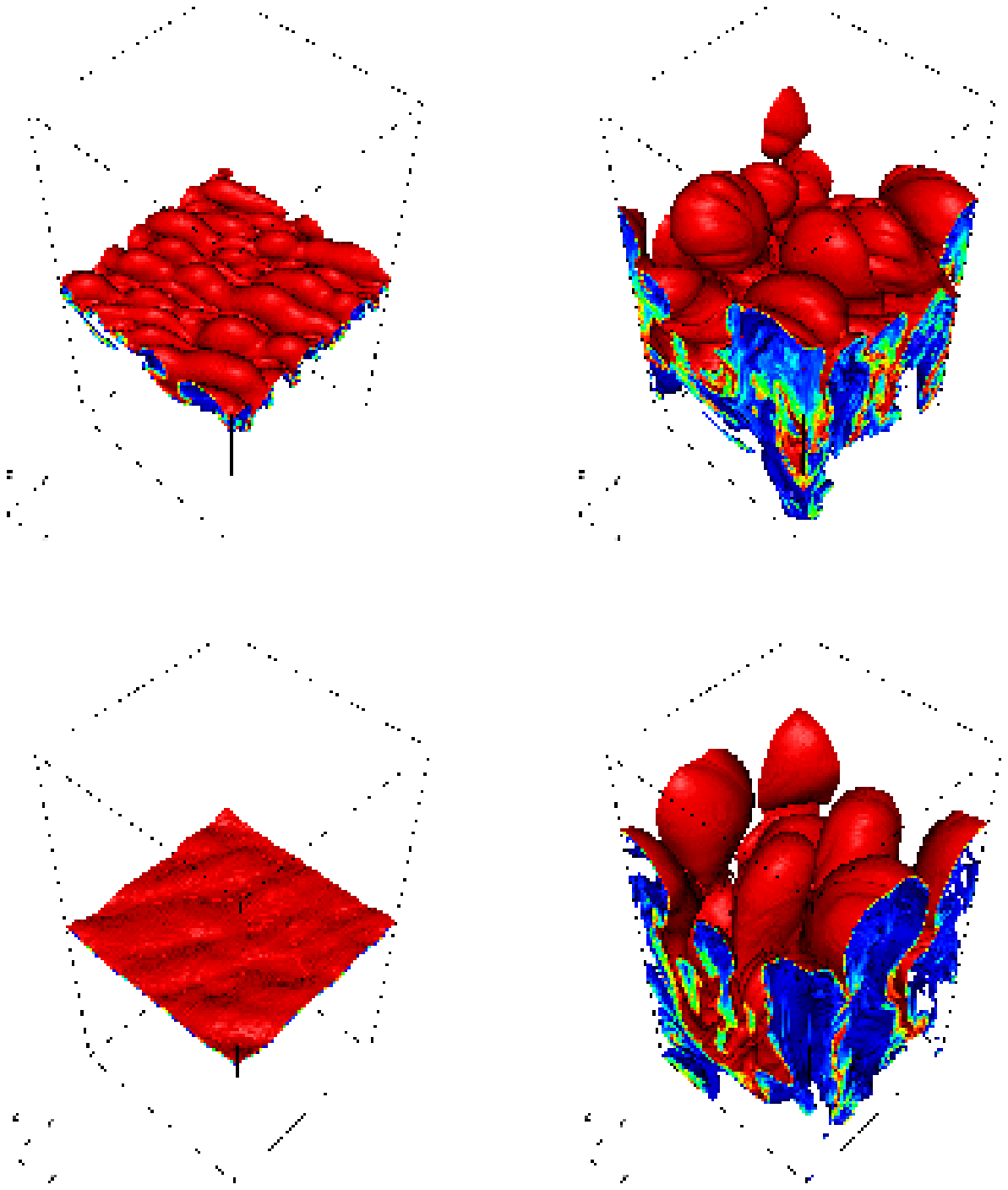}
\figcaption
{Isosurfaces of the density at $\rho=9.9$ and $\rho=1.1$ at times $t/t_s
= 20$ (left panels) and $t/t_s=40$ (top right panel) or $t/t_s=50$
(bottom right panel) in runs R45 (top, field rotated by $45^{\circ}$)
and U (bottom, field rotated by $90^{\circ}$).  Also shown are slices
of the density at the edges of the computational domain.}
\end{figure}

% Figure 5 -- Height of bubbles C90 versus U
\begin{figure}
\epsscale{0.6}
\plotone{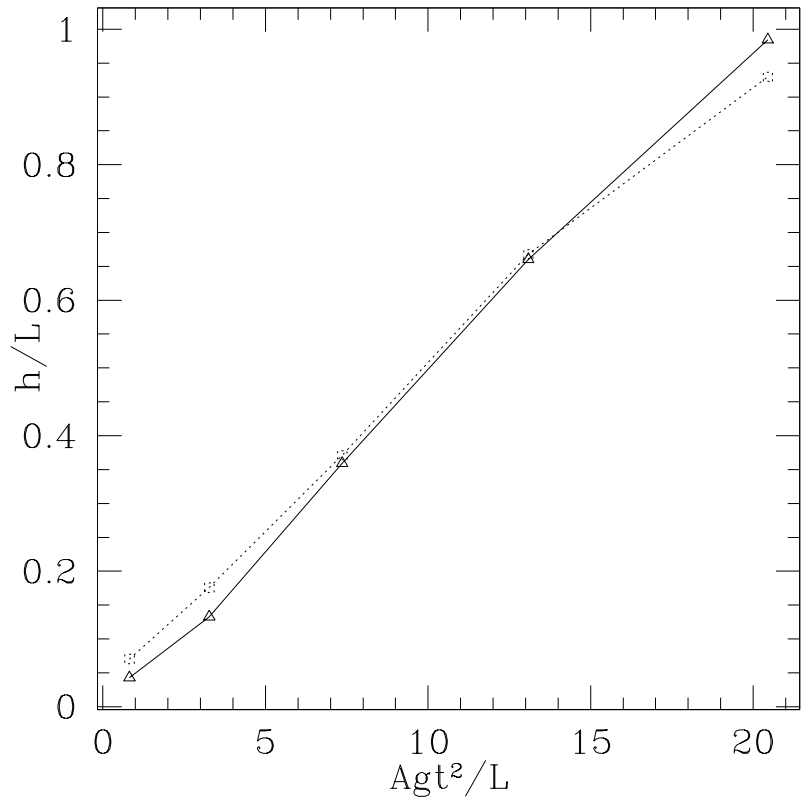}
\figcaption
{Height of bubbles as a function of time in runs U (dashed line, uniform
field) and C90 (solid line, continuous rotation over a vertical distance 
$L_{rot}/\lambda_{c} = 0.5$ at the interface).  }
\end{figure}

% Figure 6 -- Height of bubbles
\begin{figure}
\epsscale{0.6}
\plotone{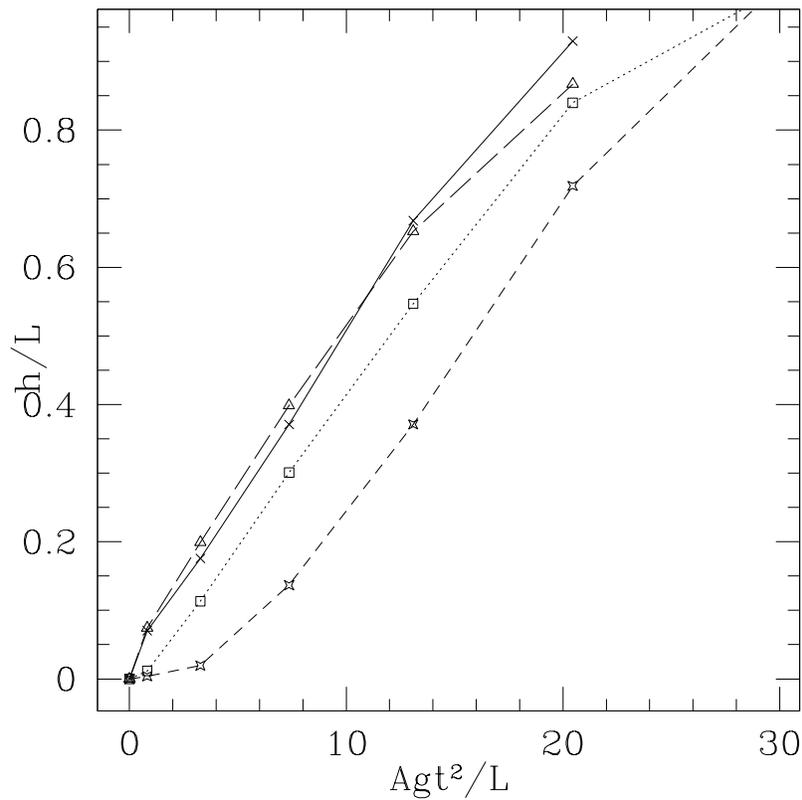}
\figcaption
{Height of bubbles as a function of time in runs U (solid line, uniform
field), LF (long dashed line, field in light fluid only), R45 (dotted
line, field rotated by $45^{\circ}$) and R90 (short dashed line, field
rotated by $90^{\circ}$).  }
\end{figure}

% Figure 7 -- Mixing parameter
\begin{figure}
\epsscale{0.6}
\plotone{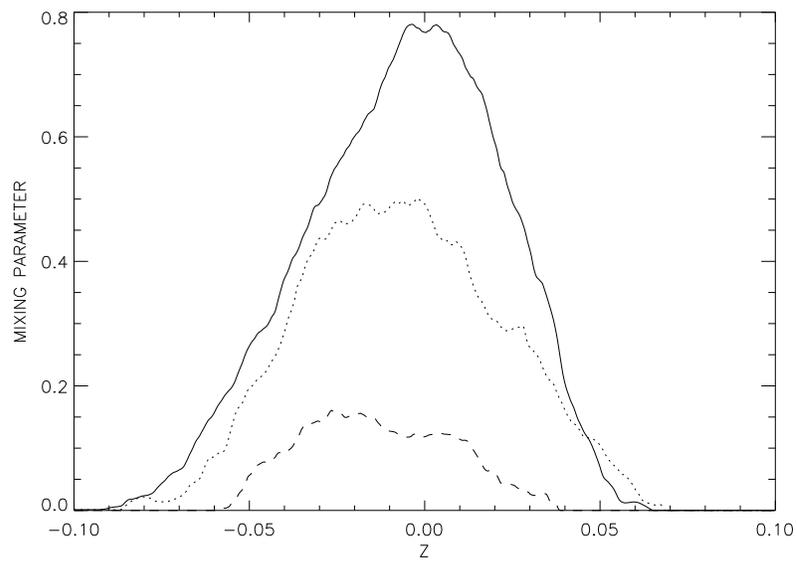}
\figcaption
{Vertical profile of the mixing parameter, defined by equation (11),
for runs H (solid line), U (dotted line), and R90 (dashed line) at 
$t/t_s=40$.  A
value of zero indicates no mixing, while one indicates fully mixed.
The magnetic field clearly suppresses mixing, especially in the rotated
field case (R90).
}
\end{figure}

% Figure 8 -- Evolution of energies
\begin{figure}
\epsscale{0.6}
\plotone{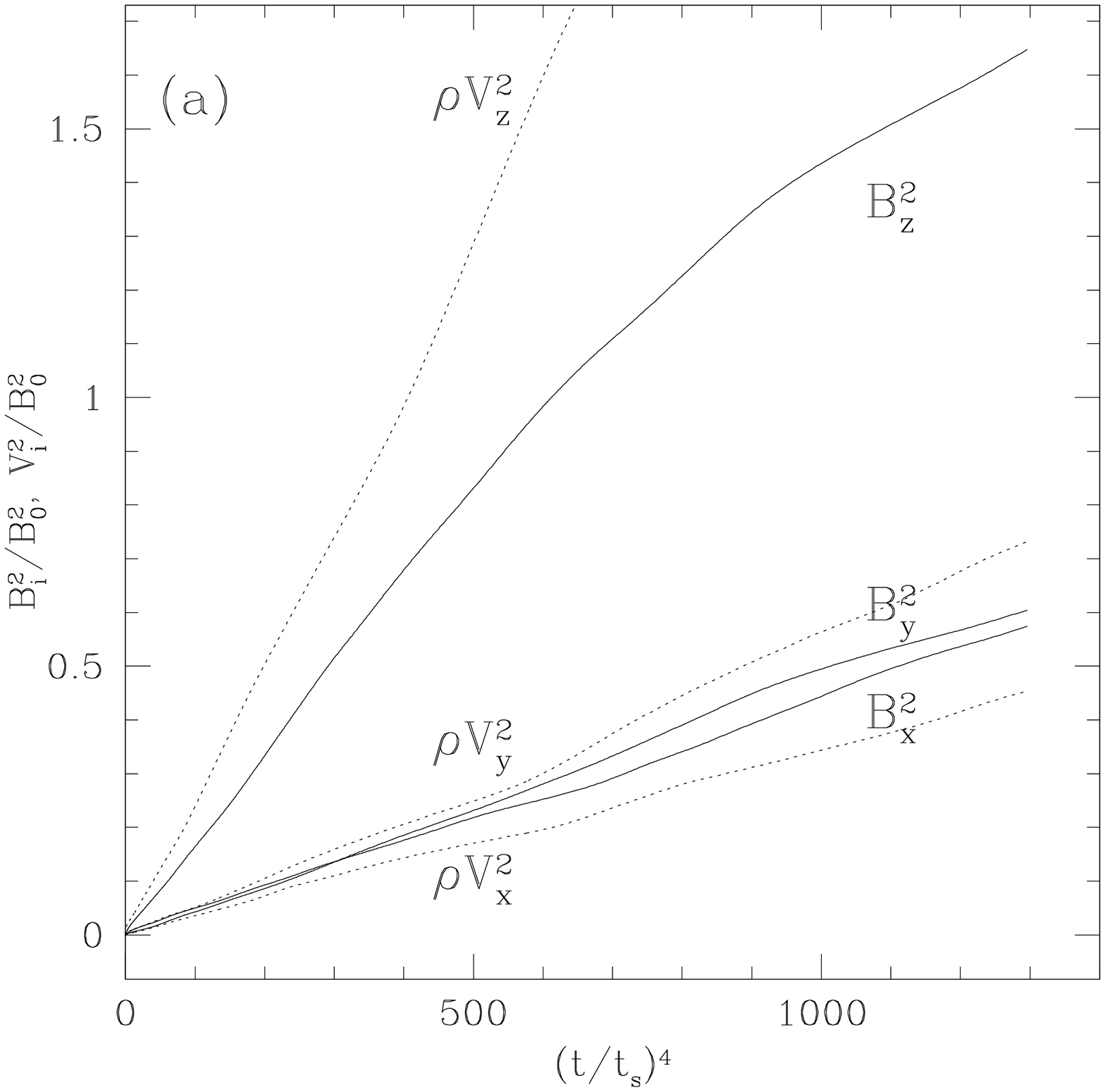}
\plotone{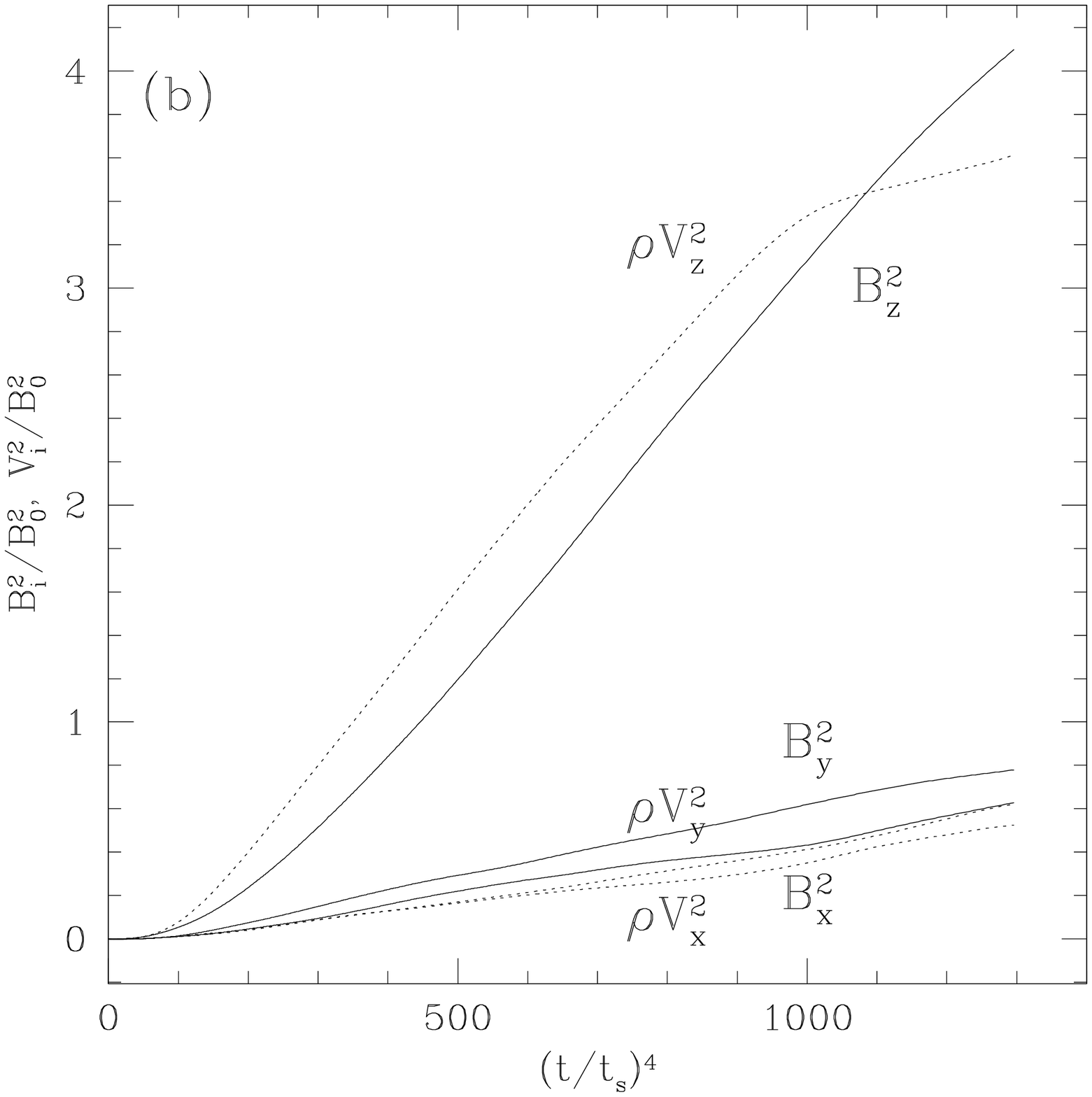}
\figcaption
{Evolution of the volume averaged kinetic and magnetic energies in
{\em (a)} run U, uniform field case, and {\em (b)} run R90, field
rotated by $90^{\circ}$ at the interface.  All components of the energy
are normalized by the volume averaged magnetic energy in the initial
conditions $B_0^{2}/2$.   In addition, the energies associated with the
$x-$ and $y-$components of the magnetic field have their initial values
subtracted as appropriate.
}
\end{figure}

% Figure 9 -- R90 upside down
\begin{figure}
\epsscale{0.6}
\plotone{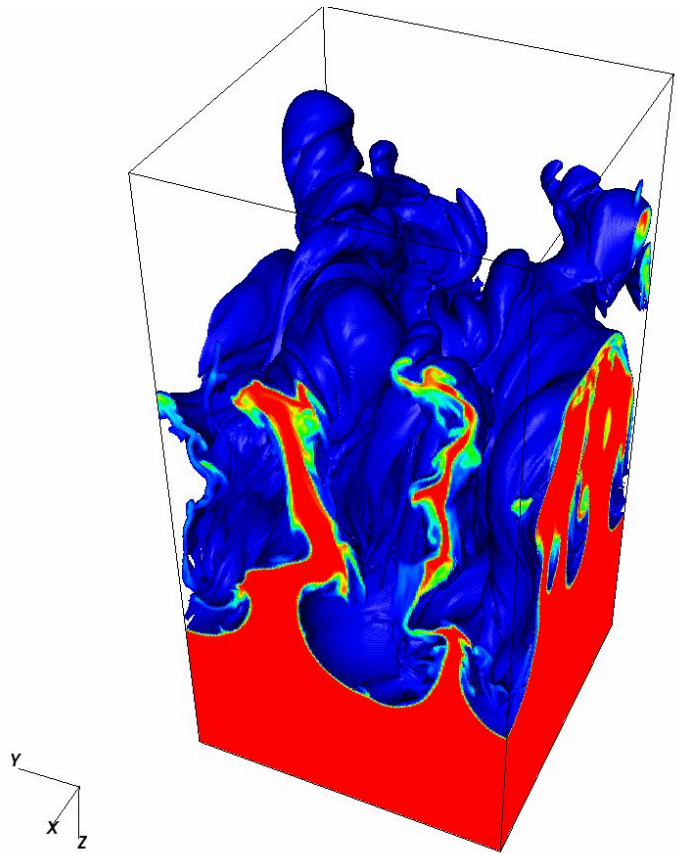}
\figcaption
{Isosurface of the density at $\rho=1.1$ at time $t/t_s=45$
in run R90 (field rotated by $90^{\circ}$).
Also shown are slices
of the density at the edges of the computational domain for $\rho>1.1$.
Compare to the lower RH panel in figure 4; noting the
orientation of the image is reversed, that is
descending fingers of heavy fluid point upward here.}
\end{figure}

\end{document}